\documentstyle[sprocl]{article}

\def\be{\begin{equation}}
\def\ee{\end{equation}}
\def\bea{\begin{eqnarray}}
\def\eea{\end{eqnarray}}

\begin{document}
\begin{flushright}
{CERN-TH/2001-203 }\\
\end{flushright}

\vspace{1cm}
\title{LARGE PHASES AND CP VIOLATION IN SUSY}

\author{TAREK IBRAHIM}

\address{Department of Physics, Faculty of Science, University
of Alexandria,\\ Alexandria, Egypt} 

\author{PRAN NATH}

\address{Theoretical Physics Division, CERN CH-1211, Geneva 23,
Switzerland\\
Department of Physics, Northeastern University,\\
 Boston, MA 02115-5000, USA\footnote{Permanent address}}


\maketitle\abstracts{ A brief overview is given of recent developments
in the analyses of large phases and CP violation in supersymmetric 
unified models. The problem of experimental electric dipole moment
constraints and large phases is discussed. Implications of large
phases on supersymmetric phenomena are reviewed. The possibility
of generating a muon electric dipole moment much larger than implied
by the scaling relation $d_{\mu}/d_e\simeq m_{\mu}/m_e$ 
from lepton flavor nonuniversality and within reach of the recently
proposed Brookhaven experiment for a sensitive probe of $d_{\mu}$
is also discussed.}


\section{Introduction}
In this paper we will give an overview of the subject of large phases
and CP violation in supersymmetric (SUSY) unified models. Specifically we 
will discuss the EDM problem in SUSY arising from 
the current experimental EDMs  constraints\cite{commins},
their satisfaction with large phases and the effect of large
phases on SUSY phenomena. We will also discuss the possibility of
 generating a muon edm (EDM) significantly
 larger than that dictated by a linear scaling in the lepton mass. 
 The muon EDM is of considerable current interest in view of a
 recent proposal for a sensitive measurement of it at 
 Brookhaven\cite{sem1}.
 We begin our discussion regarding the situation in the 
 electro-weak sector of standard model. Here there is only one 
 CP phase which arises in the Kobayashi-Maskawa mass matrix and 
 this phase contributes to the lepton EDMs only at the multiloop
 level and consequently the lepton EDMs in the Standard Model are 
 extremely small\cite{hoogeveen} and beyond the reach of current
 experiment and also beyond the reach of any conceivable experiment
 in the near future. 
  It is known that baryogenesis requires a new source of 
CP violation beyond what is in the standard model. Thus new 
CP violating phases must exist in nature beyond what is in the
SM. Such new phases would also contribute to the lepton EDMs and 
 consequently the lepton EDMs provide a very clean window for discovering
 new physics. The
QCD sector sector of the standard model is more complex as it brings in 
another phase arising from a  topological term in the effective QCD 
Lagrangian, i.e.,  $\theta_G\frac{\alpha_s}{8\pi}G\tilde G$.
The effective parameter which controls CP violation is 
$\bar\theta =\theta_G+arg (detM_uM_d)+..$ which gives a neutron EDM
$d_n\simeq 1.2\times 10^{-16}\bar \theta ecm$.  The current limit 
$ d_n<6.5\times 10^{-26} ~ecm$ implies $\bar\theta< 6\times 10^{-10}$.
The desired smallness of $\theta$ is the well known problem of QCD
which has been discussed quite extensively in the literature.
The same problem, of course, also persists in supersymmetric theories
(For a recent discussion of this problem see Ref.\cite{ma}).
 However, even beyond the $\theta$ problem in QCD
there is a CP problem unique to SUSY. We discuss this in Sec.2

\section{CP PHASES IN SUSY}
Models based on soft breaking of supersymmetry contain an abundance 
of CP violating phases. Thus, for example,
mSUGRA with CP violation depends on the parameters
 $m_0, m_{\frac{1}{2}}, A_0, \tan\beta, \theta_{\mu}, \alpha_{A_0}$
where $m_0$ is the universal scalar mass, $m_{\frac{1}{2}}$ is the 
universal gaugino mass, 
$A_0=|A_0|exp{i\alpha_{A_0}}$ is the universal trilinear coupling, 
$\tan\beta$ 
is the ratio of
the two Higgs VEVs in MSSM, and $\theta_{\mu}=Arg(\mu)$ where $\mu$ is
the Higgs mixing parameter $\mu$ (we use the sign convention of
Ref.\cite{sugra}). Thus there are two phases, i.e,,
$\alpha_{A_0}$ and $\theta_{\mu}$ that enter in  mSUGRA\cite{insugra}.
The non-universal supergravity models  and MSSM involve
 many more phases and the edms of quarks and leptons will depend on these.
Thus in MSSM the electron EDM depends on three  
independent phases $\xi_i+\theta_1 (i=1,2)$ and 
$\alpha_{A_e}+\theta_{\mu}$ where $\xi_i$ are the phases of the gauginos
masses $\tilde m_i$, i.e., $\tilde m_i=|\tilde m_i|exp(i\xi_i)$
(i=1,2) corresponding to the U(1) and SU(2) gauginos. 
The  quark  EDMs depends on 9 phases $\xi_i+\theta_{\mu} (i=1,2,3)$;
 $\alpha_k+\theta_{\mu} (k=u,d,c,s,t,b)$.The electron and the neutron 
 edms together depend on ten independent phases\cite{inmssm}.
 
In a broad class of SUSY, string and brane models we expect the
CP phases of O(1) as there is no a priori reason for it to be otherwise.
Phases of this size lead to an EDM of the electron and of the neutron 
which are significantly larger than their experimental 
lower limits. 
Possible solutions to this problem consist of choosing  
small phases\cite{ellis}, assuming 
a heavy SUSY spectrum with masses O(several) TeV\cite{na}, 
embedding the models in a left-rigth symmetric framework which suppresses
the dangerous phases\cite{bdm1}, 
and the more recently proposed mechanism of internal
cancellations\cite{insugra,inmssm,incancel}.
 There is also the possibility that the phases arise
only in the third generation and hence their contributions to the
EDM of the first generation quarks and leptons are 
suppressed.  
 The dominant contributions to the lepton EDMs arise from the 
one loop chargino ($\chi^{\pm}$) and one loop neutralino ($\chi_i^0$)
 exchanges. For the case  of the neutron EDM one has contributions
 from one loop
 chargino, neutralino and gluino ($\tilde g$)
 exchanges, and  in addition contributions
 from the two loop stop-top and sbottom-bottom exchanges.
 In certain parts of the parameter space 
two loop contributions from CP odd Higgs exchange 
may also be important\cite{chang}.
The operators that contribute are the electric dipole operator
$-\frac{i}{2} d_f \bar{\psi} \sigma_{\mu\nu} \gamma_5 \psi
F^{\mu\nu}$, the chromoelectric electric dipole operator 
$-\frac{i}{2}\tilde d^C \bar{q} \sigma_{\mu\nu} \gamma_5 T^{a} q
 G^{\mu\nu a}$,and the purely gluonic dimension six operator 
 $ -\frac{1}{6}d^G f_{\alpha\beta\gamma}
G_{\alpha\mu\rho}G_{\beta\nu}^{\rho}G_{\gamma\lambda\sigma}
\epsilon^{\mu\nu\lambda\sigma}$.
In extracting the effects of the chromoelectric and  the purely 
gluonic operators one uses the so called naive dimensional analysis 
of Georgi-Manohar\cite{gm}, i.e., 
$d^C_q=\frac{e}{4\pi} \tilde d^C_{q} \eta^C$,
$d_{n}^G=\frac{eM}{4\pi} d^G \eta^G$, 
where $\eta^C$ $\approx$ $\eta^G$ $\sim 3.4$,  
 $M$ =1.19 GeV is the chiral symmetry breaking scale.
 The neutron EDM $d_n$ is estimated using SU(6) quark model 
  $d_n=(\frac{4}{3}d_d-\frac{1}{3}d_u)$.
  Another constraint recently imposed in some analyses is the
  experimental constraint of the  EDM of atoms. For example, the
  EDM of the mercury atom is extremely accurately known\cite{hg}, i.e.,
   $d_{H_g}< 9\times 10^{-28}$ ecm. However, a theoretical
    analysis of an atomic EDM depends on the Schiff moment and involves 
 nuclear physics effects which are poorely understood. A more
 accurate understanding of the Schiff moment in terms of the
 parameters of the microscopic CP violating  SUGRA or MSSM Lagrangian
  is needed to have confidence in such an analysis. 
 If the phases are large they will affect low energy phenomena.
  Thus inclusion of CP phases will affect 
sparticle masses, decay branching ratios and 
cross-sections\cite{moretti},
 neutralino relic density and detection rates
 in dark matter detectors\cite{cin}, g-2\cite{ing}, 
 higgs system\cite{pilaftsis,pilaftsis2,demir,inhiggs,i,aa},
 trileptonic signal\cite{trilep,cptrilep},
$b\bar b$ system\cite{voloshin}, baryogenesis\cite{baryogenesis},
proton decay\cite{cpproton},
 and hadron collider phenomenology\cite{mrenna} and $e^+e^-$ 
 collider phenomenology\cite{barger}.
 The possibility that soft SUSY phases may be the origin of all CP
 violation has also been considered\cite{epsilon}.
In the following we discuss the CP effects on the neutral Higgs system,
and on g-2. We will also discuss the possibility of generating a muon
EDM which is significantly larger than what is predicted by fermion
mass scaling. 

\section{CP effects in neutral Higgs system}
Soon after the possibility of large CP phases became 
feasible\cite{insugra,inmssm,incancel} it was
pointed out that CP violation through loops would generate mixing
between the CP even and the CP odd 
sectors\cite{pilaftsis}. The CP even -CP odd 
mixing was exhibited using the stop 
exchange\cite{pilaftsis,pilaftsis2,demir}. More recently it 
was pointed out that for large $\tan\beta$ effects of chargino 
exchange would be significant and may become as large or even larger
than the stop exchange\cite{inhiggs}. We illustrate here the main elements of 
this analysis.  In the presence of large CP violating phases 
the spontaneous symmetry breaking including one loop effects 
generates an induced phase so that  
\begin{equation}
(H_1)= \left(\matrix{H_1^0\cr
 H_1^-}\right)
 =\frac{1}{\sqrt 2} 
\left(\matrix{v_1+\phi_1+i\psi_1\cr
             H_1^-}\right) 
\end{equation}	     

\begin{equation}
(H_2)= \left(\matrix{H_2^+\cr
             H_2^0}\right)
=\frac{e^{i\theta_H}}{\sqrt 2} \left(\matrix{H_2^+ \cr
             v_2+\phi_2+i\psi_2}\right)
\end{equation}
In the basis
  $\{ \phi_1,\phi_2,\psi_{1D}, \psi_{2D}\}$ where
$\psi_{1D}=\sin\beta \psi_1+ \cos\beta \psi_2$, and
$\psi_{2D}=-\cos\beta \psi_1+\sin\beta \psi_2$, 
$\psi_{2D}$ decouples and the remaining $3\times 3$ mass$^2$ matrix 
 $M^2_{Higgs}$ is given by 

\begin{equation}
\left(\matrix{M_Z^2c_{\beta}^2+M_A^2s_{\beta}^2+\Delta_{11} &
-(M_Z^2+M_A^2)s_{\beta}c_{\beta}+\Delta_{12} &\Delta_{13}\cr
-(M_Z^2+M_A^2)s_{\beta}c_{\beta}+\Delta_{12} &
M_Z^2s_{\beta}^2+M_A^2c_{\beta}^2+\Delta_{22} & \Delta_{23} \cr
\Delta_{13} & \Delta_{23} &(M_A^2+\Delta_{33})}\right) 
\end{equation}
In Ref.\cite{pilaftsis,pilaftsis2,demir} stop corrections to $m_A^2$ 
and to $\Delta_{ij}$ 
(i,j=1,2,3) were computed and it was shown that all of the 
  Q scale dependence can be absorbed in $m_A^2$ and
that $\Delta_{ij}$  are scale independent. One then finds
that the diagonalization of the mass$^2$ matrix of Eq.(3) leads to 
mixing in the mass diagonal eigenstates between the CP even and the 
CP odd
components. In Ref.\cite{inhiggs} this analysis was extended to include
the W-chargino($\chi^+$)-charged Higgs ($H^+$) exchange. 
It was shown that a composite treatment of $W-\chi^+-H^+$ 
exchange allows one to absorb all the Q dependence 
in  $m_A^2$ and the $\Delta_{ij}$ once again have no explicit
Q dependence\cite{inhiggs}. 
With inclusion of the chargino exchange contribution
$m_A^2$ now reads\cite{inhiggs}
\begin{eqnarray}
m_A^2=(\sin\beta\cos\beta)^{-1}(-m_3^2\cos\theta_H +
+\frac{g_2^2}{16\pi^2}|\tilde m_2||\mu| \cos\gamma_2 
f_1(m_{\chi_1^+}^2, m_{\chi_2^+}^2)) +.. 
\end{eqnarray}
where $f_1(x,y)=-2+log(xy/Q^4)+((y+x)/(y-x))log(y/x)$ and contains
the explicit Q dependence and ..represent the contributions from 
the stops, sbottoms etc.
The $\tilde W-W-H^+$ exchange contribution to the lightest  higgs boson
mass  is typically negative and lies in the range of 1-2 GeV 
and one needs to include this effect in the precision analyses.
$\tilde W-W-H^+$  also contributes to the CP even-CP odd Higgs mixing.
While as in previous analyses the lightest higgs
typically remains a CP even state, there is a significant mixing
between between the heavy CP even neutral Higgs boson $H^0$ and the CP odd
Higgs boson $A^0$. The relative strength  of the chargino exchange
contribution vs the stop exchange contribution depends on $\tan\beta$
and for $\tan\beta \geq 30$ the chargino 
contributions can dominate the stop contribution. 
If large CP phases exist, then CP even -CP odd Higgs mixing 
could be seen at $e^+e^-$ colliders and would
provide a clear signal for the existence of such phases\cite{aa}.
Further, it was shown in Ref.\cite{i} that if CP even-CP odd Higgs
mixing is seen experimentally then it is only the cancellation 
mechanism that can explain such a mixing\cite{i} consistent with
EDM constraints.

\section{CP Effects on g-2}
 One of the phenomena affected by SUSY CP phases is 
 the supersymmetric contribution to $g_{\mu}-2$. It was shown 
 in Ref.\cite{ing} that the supersymmetric contribution to 
 $a_{\mu}=(g_{\mu}-2)/2 $ is strongly dependent
 on the phases $\theta_{\mu}$ and $\xi_2$ and also 
 dependent, though somewhat less strongly, on the phases
 $\xi_1$ and on 
 $\alpha_{A_0}$. One may ask what the implication of this strong
 dependence is for CP phases in light of  the recent Brookhaven
 data which finds a discrepancy between experiment 
 and the Standard Model prediction such that\cite{brown}   
 $a_{\mu}^{exp}-a_{\mu}^{SM}=43(16)\times 10^{-10}$.
 This question was investigated in Ref.\cite{icn} and it is
 found that the BNL data constrains the CP phases very strongly.
Thus one finds that as much as 60-90\% of
the parameter space in the $\theta_{\mu}-\xi_2$ plane is eliminated
by the BNL constraint\cite{icn}. Further, 
 it is possible to construct models with large CP phases
which satisfy the EDM constraints as well as the Brookhaven
constraint on g-2\cite{icn,arnowittedm}. 
Five models of this type are exhibited in
Table 1 where the phases are large, EDM constraints on the electron
and on the neutron EDM are satisfied, and $a_{\mu}^{(SUSY)}$
lies in the range given by the BNL experiment.
 One also finds\cite{icn} that all of the sparticle spectrum
corresponding to Table 1 is consistent with naturalness constraints 
(see, e.g., Ref\cite{ccn}) and would be accessible at hadron colliders 
 and some of the spectra may also be accessible at linear colliders. 
\begin{table}[h]
\begin{center}
\caption{{Cases where 
the EDM and the g-2 experiments are 
satisfied (from Ref.[33])}}
\begin{tabular}{|l|l|l|}
\hline
\hline
(case) $\xi_2$, $\theta_{\mu}$, $\xi_3$ 
 & $d_e$, $d_n$ (ecm)& 
$a_{\mu}^{SUSY}$\\
\hline
\hline
(a) $-.63$,$.3$,$.37$& $-4.2\times 10^{-27}$, $-5.3\times 10^{-26}$ & $47.0 \times 10^{-10}$\\
\hline
\hline
(b)$-.85$ ,$.4$ ,$.37$
 & $4.2\times 10^{-27}$, $4.8\times 10^{-26}$ &
$10.8\times 10^{-10}$  \\
\hline
\hline
(c)$-.8$ ,$.2$ ,$1.3$
 & $4.0\times 10^{-27}$, $5.4\times 10^{-26}$ &
$12.2\times 10^{-10}$  \\
\hline
\hline
(d)$-.32$ ,$.3$ ,$-.28$
 & $-1.2\times 10^{-27}$, $3.3\times 10^{-26}$ &
$20.1\times 10^{-10}$  \\
\hline
\hline
(e)$-.5$ ,$.49$ ,$-.5$
 & $1.8\times 10^{-27}$, $-6.6\times 10^{-27}$ &
$12.7\times 10^{-10}$  \\
\hline
\hline
\end{tabular}
\end{center}
\end{table}
\section{ Large Muon EDM}
There is a recent Brookhaven proposal\cite{sem1} to  probe 
$d_{\mu}$ with a sensitivity of
 $d_{\mu}\sim O(10^{-24})ecm$. 
In most theoretical models the charge lepton edms scale, e.g.,
$\frac{d_{\mu}}{d_e}\simeq \frac{m_{\mu}}{m_e}$.
Since experimentally  $d_e< 4.3\times 10^{-27}ecm$ the scaling relation
 if valid implies that $d_{\mu}\leq 10^{-24}ecm$  which, however,
falls below the sensitivity of the proposed BNL experiment.
Thus a large muon edms can be gotten only by the breakdown of scaling.
Some models where this comes about consist of the
two higgs doublet model\cite{barger2}, left-right symmetric 
models\cite{bdm2,ng}, and
models with flavor non-universalities in the slepton 
sector\cite{inmuedm}(see also Ref.\cite{feng}).
We discuss here the last possibility, i.e., models with 
slepton flavor nonuniversality. To illustrate in some detail
how the scaling relation gets violated in this case, we consider
the charge lepton edm  arising from the exchange of  
 charginos and neutralinos which is given by\cite{inmssm}
\begin{eqnarray}
d_{\it l}=\frac{e\alpha_{EM}}{4\pi \sin^2\theta_W}
\frac{\kappa_{\it l}}{m_{\tilde \nu_{\it l}}^2}
\sum_{1=1}^{2}\tilde m_{\chi_i^+}Im(U^*_{i2}V_{i1})
A(\frac{\tilde m_{\chi_i^+}^2}{m_{\tilde \nu_{\it l}}^2})\nonumber\\
+\frac{e\alpha_{EM}}{4\pi \sin^2\theta_W}
\sum_{k=1}^{2}\sum_{i=1}^{4} Im (\eta_{ik}^{\it l})
\frac{\tilde m_{\chi_i^0}}{M_{\tilde {\it l}_k}^2}
Q_{\tilde {\it l}}B(\frac{\tilde m_{\chi_i^0}^2}{M_{\tilde {\it l}_k}^2})
\end{eqnarray}
where $\kappa_{\it l}=m_{\it l}/(\sqrt 2 m_W \cos\beta)$ and 
 $Im(\eta_{ik}^{\it l})$ is given by 
\begin{equation}
Im(\eta_{ik}^{\it l})=m_l(C_{jk}+ A_l  d_{jk}+..)
\end{equation}
We see now that if $A_l$ is universal, i.e., $A_e=A_{\mu}$, 
one has scaling for the EDMs,
$\frac{d_{\mu}}{d_e}\simeq \frac{m_{\mu}}{m_e}$. However, 
in the presence of non-universality $A_{\mu}\neq A_e$ and 
the scaling relation breaks down. In this case
slepton flavor  nonuniversality can upset the cancellation mechanism 
in the  muon EDM even when such a cancellation occurs for the
EDM of the electron. In  this situation the cancellation mechanism 
produces an EDM of the electron consistent with the current 
experimental limit while the lack of cancellation in the muon 
channel produces an muon EDM much larger than what scaling predicts
and in the range accessible to the proposed Brookhaven 
experiment.

\section*{Acknowledgments}
This research was supported in part by NSF grant PHY-9901057.


\end{document}